%% file: polychromatic.tex
\documentclass[%
 reprint,
 superscriptaddress,
 amsmath,amssymb,
 aps,
 prb,
]{revtex4-1}

\usepackage{overpic}
\usepackage{booktabs}
\usepackage{framed}
\usepackage{comment}
\usepackage{siunitx}
\include{new_preamble_definitions}
\newcommand*\pkg[1]{\textsc{#1}}
\begin{document}

\title{A Framework for Formulating Polychromatic Theories of Emission}
\author{Ivan Fernandez-Corbaton}
\email{ivan.fernandez-corbaton@kit.edu}
\affiliation{Institute of Nanotechnology, Karlsruhe Institute of Technology, Kaiserstr. 12, 76131 Karlsruhe, Germany}
\author{Maxim Vavilin}
\affiliation{Institut f\"ur Theoretische Festk\"orperphysik, Karlsruhe Institute of Technology, Kaiserstr. 12, 76131 Karlsruhe, Germany}
\author{Markus Nyman}
\affiliation{Institute of Nanotechnology, Karlsruhe Institute of Technology, Kaiserstr. 12, 76131 Karlsruhe, Germany}
\begin{abstract}
	The emission of energy as electromagnetic radiation is ubiquitous, in particular because objects release thermal energy in the form of photons. Most theories of thermal radiation assume that the thermal emissions originate from a continuum of elementary monochromatic sources, uncorrelated to each other. The universality of thermal radiation motivates the consideration of theories that allow for more general kinds of elementary emissions.
	In here, we introduce a framework for formulating polychromatic theories of emission in the electromagnetic Hilbert space, whose computational side is based on the transition matrix, or T-matrix. Each photon is emitted as a coherent polychromatic pulse. The spectra of the different emitted pulses are derived using the natural resonance frequencies of the given finite-size object. Each resonance belongs to one of the orthogonal subspaces which decompose the absorption operator according to the symmetries of the object. Energy conservation in the steady-state is ensured by equalizing the absorption and emission of energy at each individual subspace. The framework can accommodate general illuminations, and produce emissions with frequencies that are much suppressed in or even absent from the illumination, resulting in different rates of emission and absorption of photons. This makes the framework suitable for describing other kinds of emissions, such as luminescence, in the Hilbert space.
\end{abstract}
\maketitle
In the works of Kirchhoff and Planck \cite{Planck1914,Kirchhoff1921}, the thermal equilibrium radiation is isotropic, unpolarized, and uncorrelated in direction, time, and frequency. Both of them used ray optics and assumed the objects under consideration and their features to be much larger than the wavelength \cite{Baltes1976}\cite[Chap.~4.7]{Bohren1983}. Such assumption is grossly violated when modern fabrication techniques are used to structure matter at the micrometer scale. Then, measurements have already shown that thermal radiation can be polarized \cite{Krueger2011,Oehman1961}, and exhibit directionality \cite{Greffet2002} and time correlations \cite{Greffet2007,Babuty2013}. 

Most theories of thermal emission are either based on fluctuation electrodynamics (FE) \cite{Rytov1959,Kattawar1970,Eckhardt1984}, or on the Kirchhoff law of thermal radiation \cite{Kirchhoff1921,Miller2017,Guo2022}. Both bases allow one to predict thermal radiation that is polarized, and correlated in direction and also in time. In contrast, emissions are assumed to originate from independent monochromatic sources, which implies the absence of correlations between different frequencies of the emitted spectra. In FE, this results from assuming that the random current source with frequency $\omega$ a given point $\rr$ inside the object has no relationship with the random current source at the same point $\rr$ but at a different frequency $\bar{\omega}$, independently of the distance between the two frequencies, as formalized with a delta distribution $\delta(\omega-\bar{\omega})$. 

The assumption of monochromatic sources leading to a frequency-uncorrelated emission spectrum originates in the gedanken experiments involving thermal fields inside a cavity with perfectly reflecting walls. In this case, the cavity modes are indeed monochromatic and independent, and such a property was then assumed for the thermal radiation escaping through a small hole in the cavity \cite[p.~43]{Planck1914}. In more realistic systems, the modes have complex frequencies, which imply extended polychromatic spectra on the real frequency axis. Assuming that each mode emits single photons with extended spectra leads to frequency correlations. Such correlations can be investigated by interferometry, or coincidence counts between different frequency components \cite{Silva2016,Ferrantini2025}. 

There is ongoing work on more nuanced theories of thermal radiation \cite{Morino2017,Greffet2018,Miller2017}. Due to the universality of thermal radiation, refinements in its theoretical understanding are relevant in many areas of physics. In particular, changes in the theory will affect the performance upper bounds for radiative transfer of heat \cite{Miller2015}, and the engineering of thermal emissions \cite{Manjavacas2014,Yu2017,Picardi2023,Hamdan2022} for applications such as thermo-photovoltaic energy conversion and daytime radiative cooling \cite{Fan2022,Xin2023}. However, as far as we know, only the approach developed for spheres in \cite{Morino2017} has proposed polychromatic emission modes. It should also be noted that there is often discrepancies between experimental results and the predictions of even the most advanced theories \cite[Sec.~IV]{Song2015}. Such discrepancies can be partly attributed to experimental uncertainties, and to the need of refined models for the optical response of hot materials \cite{Fenollosa2019}, but it is not clear whether those are the only contributing factors. These considerations motivate the formulation of polychromatic theories of thermal emission and, moreover, highlight the value of a unified framework capable of describing diverse emission processes, including luminescence. We note that there already exist theories that establish connections between thermal radiation and luminescence \cite{Wurfel1982,Greffet2018}.

We will work in an electromagnetic Hilbert space \cite{Gross1964,Zeldovich1965,Birula1996,FerCor2024b} which allows one, in particular, to readily formulate and compute the number of photons of a given field, and the absorption operator of a particular object. Additionally, the random character of thermal fields is readily included in the formalism. The computations are facilitated by the natural connection between the Hilbert space $\M$ and the transition matrix, or T-matrix. The T-matrix \cite{Waterman1965} is a popular and powerful approach to computations in light-matter interactions \cite{Gouesbet2019,Mishchenko2020,Wriedt2008a,Hellmers2009}, that can be applied to macroscopic objects and also to molecules \cite{FerCor2018,Zerulla2022}. The combination of $\M$ and T has been recently used in a methodology for computing thermal radiation using the directional Kirchhoff law, applicable to nano-particles, clusters thereof, and also molecules \cite{Mazo2025}.

In here, we introduce a framework to formulate polychromatic emission theories, where each photon is emitted as a coherent polychromatic pulse of light. In Sec.~\ref{sec:newdecom}, we decompose the absorption by a given object into orthogonal subspaces, and then define radiating polychromatic modes corresponding to each subspace. The pulses are derived using the natural resonances of the system by enforcing that such pulses can be normalized to contain one photon. We discuss their conceptual advantages over the typical Lorentzian lineshapes. The total emission is the sum of temporal trains of single photon emissions through each mode, with random delays and random phases whose statistics can be selected. The study of how such selections control the statistics of the total emission, such as e.g. ``bunching'', is left for future work. As an exemplary use of the framework, we formulate in Sec.~\ref{sec:model} a simple polychromatic theory of thermal radiation and apply it to a SiC sphere. The resulting radiation is compared to the one obtained with the Kirchhoff law. While the results are not fully satisfactory, in particular because of the high-frequency tail of the predicted spectra, the example illustrates important properties of polychromatic emission theories: emission in frequencies that the illumination does not contain, and different photon emission and absorption rates. In Sec.~\ref{sec:properties} we discuss these properties, and explain how the framework can easily handle any external illumination, such as a laser beam. These characteristics make the framework suitable for describing luminescence. Section~\ref{sec:outlook} provides a brief outlook and concludes the article.

\section{The framework in the Hilbert space\label{sec:newdecom}}
The mathematical setting that we use in this article is based on the Hilbert space of solutions of Maxwell equations, $\mathbb{M}$. We use the bra--ket notation by Dirac, and denote a particular solution of Maxwell equations as $\ket{f}$. Such setting has been recently reviewed \cite{FerCor2024b}, and appendix~\ref{app:conventions} contains a very brief overview thereof, including  expressions for the scalar product between two fields $\braket{f|h}$. The projection of a field $\ket{f}$ onto itself, $\mybraket{f}{f}$, is equal to the number of photons in the field \cite{Gross1964,Zeldovich1965}. Also, the amount of fundamental quantities such as energy and momentum in a given field can be conveniently expressed as $\braket{f|\Gamma|f}$, where $\Gamma$ is the self-adjoint operator representing a given quantity: 
\begin{equation}
	\label{eq:tableA}
		\text{Number of photons } =\braket{f|f}, \ \text{Energy } =\braket{f|\op{H}|f}.
\end{equation}
The action of the energy operator $\op{H}$ on monochromatic fields is their multiplication by $\hbar\cz k$, where $\hbar$ is the reduced Planck constant, $\cz$ the speed of light, and $k$ the angular wavenumber.

The linear light-matter interaction is modeled by means of a linear operator $S$, which depends on the material system at hand, and maps a given incoming illumination $\ket{f}$ onto the total outgoing field $\ket{g}$ resulting from the interaction: 
\begin{equation}
		\text{Linear light-matter interaction: } \ket{g}=\op{S}\ket{f}.
\end{equation}

\begin{figure*}[ht!]
		\includegraphics[width=\linewidth]{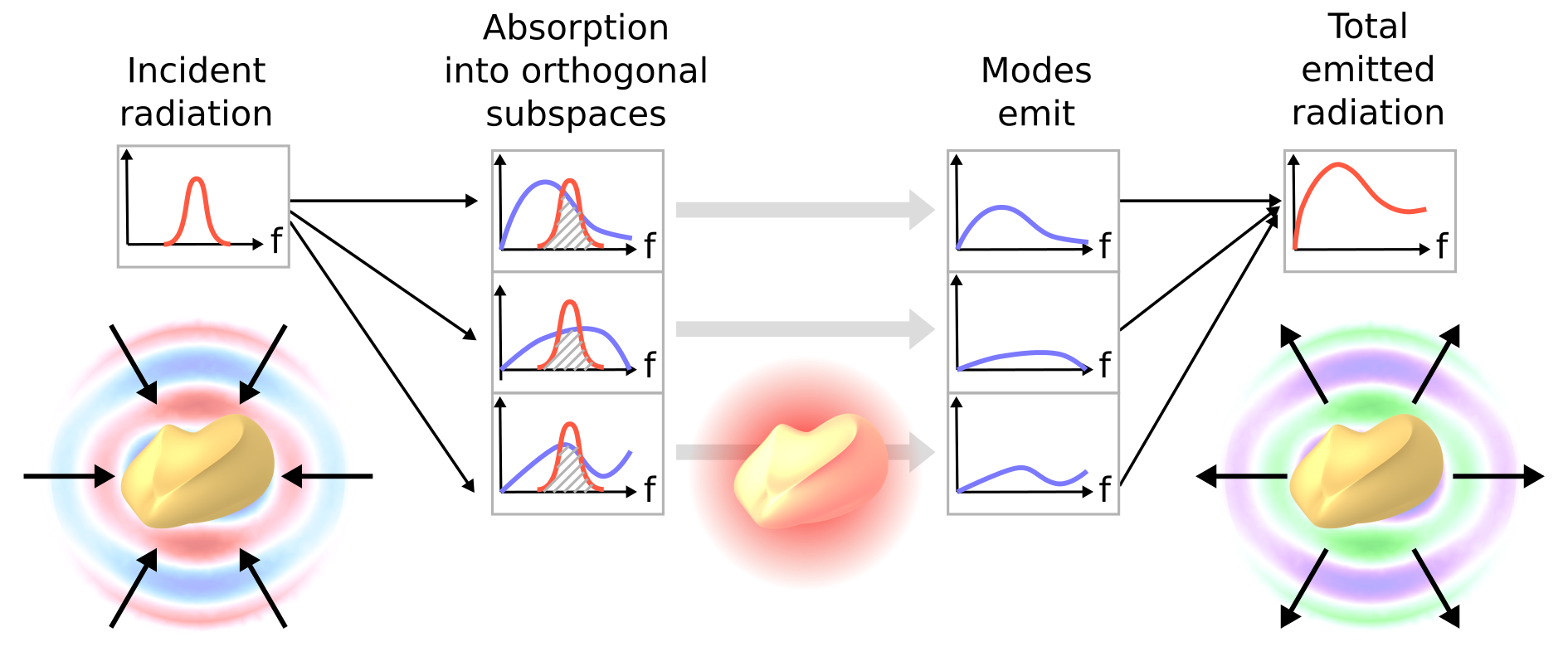}
		\caption{\label{fig:concept} Conceptual illustration of the polychromatic framework for electromagnetic emission. An incident field with a known spectrum is decomposed into the orthogonal absorption subspaces of a given object, each of which absorbs a fraction of the incident power. The energy absorbed into each subspace is re-emitted through polychromatic modes connected to the same subspace, which are obtained using the natural resonance (complex) frequencies of the object. The total emission spectrum is produced by the combination of all the polychromatic emissions.}
\end{figure*}

The change in fundamental quantities such as number of photons and energy can then be written as:
\begin{equation}
	\label{eq:tableB}
	\begin{split}
		\text{Absorbed photons }&=\braket{f|f}-\braket{g|g}=\sandwich{f}{\Id-\op{S}^\dagger \op{S}}{f}\\
								&=\sandwich{f}{\op{Q}}{f}\\
		\text{Absorbed energy}=&\braket{f|\op{H}|f}-\braket{g|\op{H}|g}=\sandwich{f}{\op{H}-\op{S}^\dagger \op{H}\op{S}}{f},
	\end{split}
\end{equation}
where $\op{X}^\dagger$ denotes the Hermitian conjugate of $\op{X}$, and the second line in \Eq{eq:tableB} defines the absorption operator $\op{Q}$.

Figure~\ref{fig:concept} is a pictorial illustration of the polychromatic framework for electromagnetic emission that we will now build.

\subsection{Decomposition of the absorption into orthogonal subspaces}
The $\op{T}$ operator is introduced by its relationship with $\op{S}$: $\op{S}=\Id+\op{T}$. In principle, $\op{T}$ can be obtained for any finite object. The absorption operator can be written as:
\begin{equation}
	\begin{split}
		\bra{f}\op{Q}\ket{f}=\bra{f}-\op{T} - \op{T}^\dagger - \op{T}^\dagger \op{T}\ket{f}.
	\end{split}
\end{equation}

From now on, we will assume that the $\op{T}$ operator is diagonal in frequency, that is, that the scattering from the object does not couple different frequencies. In such case, $\op{Q}$ can be written in the basis of multipolar fields as the following integral in angular wavenumber $k$:
\begin{equation}
	\label{eq:qkknew}
		\op{Q}=\sum_{jm\lambda}\sum_{\jbar\mbar\barlambda}\intdkmeasure \ \op{Q}^{jm\lambda}_{\jbar\mbar\barlambda} (k) |k j m \lambda\rangle\langle \barlambda\mbar\jbar k|,
\end{equation}
where the triplets $(jm\lambda)$ and $(\jbar\mbar\barlambda)$ label the rows and the columns of the monochromatic absorption matrix $\matr{{\op{Q}}}(k)$, respectively. The label $j=1,2,...$ is the multipolar degree with $j=1$ corresponding to dipoles, $j=2$ to quadrupoles, and so on, $m=-j,-j+1,\ldots, j$ is the component of the angular momentum in the $z$ direction, and $\lambda=\pm 1$ is the helicity, or polarization handedness. The expression of $\ket{kjm\lambda}$ can be found in \Eq{eq:mpdef}.

The range of integration in \Eq{eq:qkknew}, and throughout the article, excludes the potential contribution of static fields at the $k=0$ point, whose effects are not relevant for our purposes. Static fields do not belong to the Hilbert space $\M$, which contains the dynamic electromagnetic fields \cite{Gross1964}.

With typical conventions\cite{Mishchenko1996} for the monochromatic T-matrices $\matr{\op{\tilde{T}}}$, also adopted in the publicly available software \pkg{treams}\cite{Beutel2023b}, we can obtain ${\matr{\op{Q}}}(k)$ from $\matr{\op{\tilde{T}}}(k)$ as:  
\begin{equation}
	\label{eq:qt}
	{\matr{\op{Q}}}(k)=-2{\matr{\op{\tilde{T}}}}(k)-2\matr{\op{\tilde{T}}}^\dagger(k)-4\matr{\op{\tilde{T}}}^\dagger(k)\matr{\op{\tilde{T}}}(k).
\end{equation}
Then, the singular value decomposition (SVD) can be used at each $k$ to write:
\begin{equation}
	\label{eq:svdqknew}
	\matr{{\op{Q}}}(k)=\sum_s q_s^2(k) \vec{v}_s(k)\vec{v}_s^\dagger(k)\text{, with } \vec{v}_s^\dagger(k)\vec{v}_{\bar{s}}(k)=\delta_{s\bar{s}},
\end{equation}
where the singular values meet $q_s^2(k)\ge 0$, and $\delta_{s\bar{s}}$ is the Kronecker delta. This orthogonality of the singular vectors $\vec{v}_s(k)$ is the basis for the orthogonal decomposition of the absorption operator.

We now plug \Eq{eq:svdqknew} into \Eq{eq:qkknew} and obtain the sought after decomposition of $\op{Q}$:
\begin{equation}
	\begin{split}
	\label{eq:qsk}
		\op{Q}&=\sum_{jm\lambda}\sum_{\jbar\mbar\barlambda}\intdkmeasure \\
		&\sum_s q_s^2(k) \vec{v}_s^{jm\lambda}(k){\vec{v}_s^{j'm'\lambda'}}^*(k)|k j m \lambda\rangle\langle \lambda' m' j' k|\\
		&=\sum_s \intdkmeasure q_s^2(k) \ketsabs \brasabs=\sum_s\op{P}_s,
	\end{split}
\end{equation}
where we have implicitly defined\footnote{In typical SVD implementations, the singular vectors are ordered by the non-decreasing singular values. This means that it is necessary to track the singular vectors over $k$, because crossing between the singular values may occur as the angular wavenumber changes. By assuming adiabatic variations with respect to the angular wavenumber, each singular vector corresponding to an absorption matrix calculated at a discrete angular wavenumber $k_{i+1}$ can be identified with a singular vector at $k_i$ by maximizing the inner product between both sets of vectors.}:
\begin{equation}
	\label{eq:ks}
	\ketsabs=\sum_{jm\lambda}\vec{v}_s^{jm\lambda}(k)\ket{kjm\lambda}^{\text{in}},
\end{equation}
where we explicitly indicate that, in the case of absorption, the multipolar fields are the ones meeting incoming boundary conditions.

The $\ketsabs$ inherit the orthogonality properties of the $\vec{v}_s(k)$, and, at each frequency, form an orthogonal basis for the absorption. Therefore the absorption of any illumination $\ket{f}$ is separated by the projectors $\op{P}_s$ into orthogonal subspaces:
\begin{equation}
	\label{eq:fQf}
	\begin{split}
		\bra{f}\op{Q}\ket{f}&=\sum_s \braket{f|\op{P}_s|f}=\\
		\sum_s \intdkmeasure q_s^2(k)|\brasabs f\rangle|^2&=\sum_s \intdkmeasure q_s^2(k)|f_s(k)|^2,
	\end{split}
\end{equation}
where $|f_s(k)|^2$ is defined.

Let us now consider the total energy absorption $\tau^{\text{abs}}$ using \Eq{eq:tableB} and the assumption that $\op{T}$ is diagonal in frequency, which implies that $\op{S}$ and $\op{H}$ commute, to write \cite[Eq.~(11)]{FerCor2016c}: 
\begin{equation}
	\label{eq:fQf}
\tauabs=\sandwich{f}{\op{H}-\op{S}^\dagger \op{H}\op{S}}{f}=\braket{f|\op{Q}\op{H}|f}.
\end{equation}

Using that $\op{H}\ket{ks}=\hbar\cz k\ket{ks}$, we can then write \Eq{eq:fQf} as:
\begin{equation}
	\label{eq:tausabs}
	\tauabs=\sum_s \intdkmeasure (\hbar\cz k) q_s^2(k)|f_s(k)|^2=\sum_s \tausabs.
\end{equation}

The $\op{P}_s$ projectors separate the absorption according to the symmetries of the object, which establish selection rules. For example, for an achiral sphere, each subspace $s$ is characterized by three numbers: The total angular momentum $j$, the angular momentum along one axis $m$, and parity $r$, which corresponds to the TE/TM character. The sphere, however, has several different natural resonances for each given set of values of $(jmr)$ \cite[p.~554]{Stratton1941}.

\subsection{Polychromatic emission modes\label{sec:pmodes}}
Given a finite object, we intend to derive its polychromatic emission modes using the natural resonances of the object. After the decomposition of the absorption into orthogonal subspaces, such resonances can be found as the poles of the $q_s^2(k)$ in the complex plane. Appendix~\ref{app:qfroms} contains the derivation of the following expansion of $q_s^2(k)$ into partial fractions, where we generalize results contained in \cite{Grigoriev2013} from spherical to general shapes, and use that absorption tends to zero as $k\rightarrow\infty$: 
\begin{equation}
	\label{eq:wspup}
	\begin{split}
		&-q_s^2(k)=\\
		&\sum_p \frac{q_{sp}}{\cz k-\wsp}-\frac{q_{sp}^*}{\cz k+\wspstar}+\frac{q_{sp}^*}{\cz k-\wspstar}-\frac{q_{sp}}{\cz k+\wsp}=\\
		&\sum_p \frac{2\ii\text{Im}\left\{q_{sp}\right\}\cz k+2\text{Re}\left\{q_{sp}\wspstar\right\}}{\left(\cz k-\wsp\right)\left(\cz k +\wspstar\right)}+\text{ c.c.. }	
	\end{split}
\end{equation}
The $\wsp$ are complex frequencies with positive real part and negative imaginary part. Non-linear least squares fitting strategies such as the trust region reflective method \cite{Conn2000} can be used to obtain the $q_{sp}$ and $\wsp$ from the available $q_s^2(k)$. Alternatively, methods that work in the complex frequency plane can be used, as we do in Sec.~\ref{sec:model}.

The expansion in \Eq{eq:wspup} will later determine the $k$-dependence of the polychromatic emission modes. The spatial dependence of such modes at each $k$ is determined by considering the operator:
\begin{equation}
	\op{E}=\Id-\op{S}\op{S}^\dagger,
\end{equation}
whose monochromatic components are called emissivity matrices in \cite[Eq.~(18)]{Miller2017}. The SVD of such monochromatic components:
\begin{equation}
	\label{eq:E}
	\begin{split}
		\matr{{\op{E}}}(k)&=-2{\matr{\op{\tilde{T}}}}(k)-2\matr{\op{\tilde{T}}}^\dagger(k)-4\matr{\op{\tilde{T}}}(k)\matr{\op{\tilde{T}}}^\dagger(k)\\
		&=\sum_s e_s^2(k) \vec{w}_s(k)\vec{w}_s^\dagger(k),
	\end{split}
\end{equation}
leads to:
\begin{equation}
\ketsem=\sum_{jm\lambda}\vec{w}_s^{jm\lambda}(k)\ket{kjm\lambda}^{\text{out}},
\end{equation}
where the multipolar fields meet outgoing boundary conditions. At each $k$, the $\ketsem$ are an orthogonal basis for the emission.

Considering the SVD decomposition of the scattering matrix:
\begin{equation}
	\label{eq:svdsk}
	\matr{{\op{S}}}(k)=\sum_s c_s(k) \vec{a}_s(k)\vec{b}_s^\dagger(k),
\end{equation}
and the relations
\begin{equation}
		\matr{{\op{Q}}}(k)=\Id-\matr{{\op{S}}}^\dagger(k)\matr{{\op{S}}}(k),\ \matr{{\op{E}}}(k)=\Id-\matr{{\op{S}}}(k)\matr{{\op{S}}}^\dagger(k),
\end{equation}
it is straightforward to show that:
\begin{equation}
	\begin{split}
		&\vec{w}_s(k)=\vec{a}_s(k),\ \vec{v}_s(k)=\vec{b}_s(k),\\&\text{and }e_s^2(k)=q_s^2(k)=1-|c_s(k)|^2,
	\end{split}
\end{equation}
which, in particular, implies that the expansion in \Eq{eq:wspup} also applies to $e_s^2(k)$. The incoming character of $\ketsabs$ and the outgoing character of $\ketsem$ can also be deduced from the fact that $\op{S}$ maps incoming fields onto outgoing fields.

To achieve the energy conservation in steady-state situations, all the energy absorbed through the $\ketsabs$ into the $s$-th subspace is to be re-emitted through the $\ketsem$. In this way, the selection rules due to the symmetries of the object are enforced in essentially the same way as achieved by the projectors $\vec{a}_s(k)\vec{b}_s^\dagger(k)$ in \Eq{eq:svdsk}.

We will now obtain the $k$-dependence of the polychromatic emission modes from \Eq{eq:wspup}, and the requirement that they must be normalizable. For each value of $p$, one can identify one normalizable mode, which must be built using only the resonance frequencies with negative imaginary part, $\wsp$ and $-\wspstar$, which correspond to decay in time. The other two resonance frequencies correspond to growth in time, and can therefore not be used for building physically meaningful emission modes. Appendix~\ref{app:spn} contains the derivation of the normalized emission mode:
\begin{equation}
	\label{eq:sp2}
	\begin{split}
		\ketspem&=\sqrt{\frac{2\wspre\wspim}{\arctan(\wspre/\wspim)}}\times
		\\&\intdkmeasure \frac{\ii\cz}{\left(\cz k-\wsp\right)\left(\cz k +\wspstar\right)}\ketsem,
		\\&=\intdkmeasure \alpha_{sp}(\cz k)\ketsem,
	\end{split}
\end{equation}
which meets $\braspem sp\rangle^{\text{em}}=1$, and is hence a single photon mode. The inclusion of both $\wsp$ and $-\wspstar$, and the $\ii$ factor ensure that the time dependence of the corresponding electromagnetic field can be expressed as a real-valued pulse [\Eq{eq:time}].

Since the modes defined in \Eq{eq:sp2} are built using poles with negative imaginary part, causality dictates for them that they have a starting time \cite[Sec.~III]{VanKampen1953}, that is, there exist some time $t_{\text{start}}$ such that the field is zero everywhere for any time $t<t_{\text{start}}$. 

The energy contained in each $\ketspem$ photon, is shown in Appendix~\ref{app:spn} to be:
\begin{equation}
	\label{eq:spHsp}
	\Hsp=\braspem\op{H}\ketspem=\hbar \wspre\frac{-\pi/2}{\arctan(\wspre/\wspim)}.
\end{equation}
Equation~(\ref{eq:spHsp}) has the following reassuring property. Recalling that $\wspim <0$, we see that as $\wspim\rightarrow 0$, the energy tends to $\hbar\wspre$, as it is expected for a quasi-monochromatic field with frequency $\wspre$.

Modeling emissions with the $\ketspem$ modes has some conceptual advantages over the typical Lorentzian assumption, which we now explain. The Lorentzian lineshape for the spectral photon density reads:
\begin{equation}
	\label{eq:lorentzian}
	L(k)=\frac{-\wspim}{\pi|\cz k-\wsp|^2}=\frac{-\wspim }{\pi\left[\left(\cz k-\wspre\right)^2+(\wspim)^2\right]}.
\end{equation}

\begin{figure}[th!]
	\includegraphics[width=\linewidth]{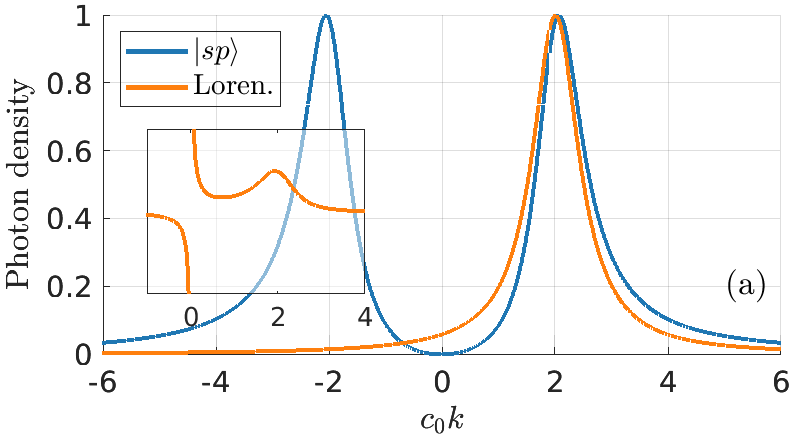}\\
	\includegraphics[width=\linewidth]{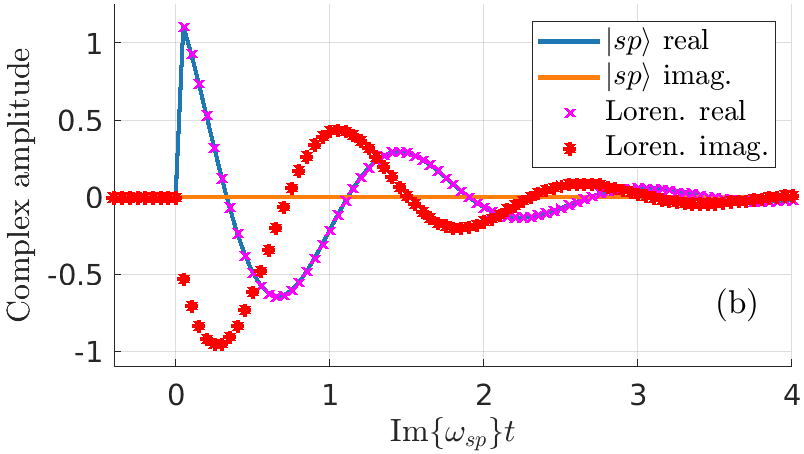}
	\caption{\label{fig:lines}Spectral photon density (a) and time dependence (b) corresponding to the $\ketspem$ modes and the common Lorentzian, as defined in \Eq{eq:sp2} and \Eq{eq:lorentzian}, respectively. The value $\wsp=2-\ii0.5$ is taken. The Lorentzian spectrum is asymmetric regarding $k\rightarrow -k$, while for $\ketspem$, all the information is already contained in $k>0$. As a consequence, the time dependence obtained in (b) is complex for the Lorentzian and real for the $\ketspem$ states [\Eq{eq:time}]. The latter is the behavior expected for physical electromagnetic fields. The total emitted energy for the Lorentzian, obtained by integrating $(\hbar\cz k)L(k)$ diverges. While this divergence can be avoided by assuming the Lorentzian lineshape for the energy density, this then causes the corresponding photon density to diverge due to the behavior of the integrand as $|k|\rightarrow 0$, as seen in the inset in (a).} 
\end{figure}
Figure~\ref{fig:lines}(a) shows both $L(k)$ and the spectral photon density\footnote{The spectral photon density of $\ketspem$ is the integrand of the $\int_{>0}^\infty\text{d}k$ integral that defines the number of photons ${}^\text{em}\langle ps\ketspem$ using \Eq{eq:amsp}.} $k|\alpha_{sp}(k)|^2$, where the two lines have been normalized to have the same maximum value. Since the Lorentzian is built from a single pole, its spectral content at $k$ is different from that at $-k$. This is an inconsistency, since the two sides of the electromagnetic spectrum should contain the same information \cite[\S~3]{Birula1996}. In particular, this asymmetry prevents one from obtaining a real-valued time dependence for the emitted electromagnetic field. In contrast, $\alpha_{sp}(k)$ contains both $\wsp$ and $-\wspstar$, all the information is already contained in $k>0$, and the emitted electromagnetic field can be expressed as a real-valued field. This is illustrated in Fig.~\ref{fig:lines}(b), which shows the two waveforms, obtained by two-sided Fourier transforms of $k\alpha_{sp}(k)$, and $l(k)=\frac{A}{k -\wsp}$. Choosing $A=\frac{\ii\exp(\ii2\psi_{sp})}{1+\exp(\ii2\psi_{sp})}$, where $\psi_{sp}$ is the phase of $\wsp$, facilitates the comparison, since then $k\alpha_{sp}(k)=l( k)+l^*(- k)$, and we obtain:
\begin{equation}
	\label{eq:time}
	\begin{split}
		\frac{A}{\cz k -\wsp}&\rightarrow-\ii \sqrt{2\pi}A\exp(-\ii\wspre t)\exp(\wspim t)u(t),\\
		\cz k\alpha_{sp}(k)&\rightarrow \text{Re}\left\{-\ii \sqrt{2\pi}A\exp(-\ii\wspre t)\right\}\exp(\wspim t)u(t),
	\end{split}
\end{equation}
where $u(t)$ is the Heaviside step function.

Importantly, the integral of the energy density $(\hbar\cz k)L(k)$ diverges because of its behavior as $|k|\rightarrow \infty$, so the total emitted energy is not well-defined. It is possible to obtain a convergent energy integral equal to $E$ if $L(k)E$ is taken as the energy density instead. However, then the corresponding photon density $L(k)E/(\hbar\cz k)$ will diverge because of its behavior as $|k|\rightarrow 0$, as shown in the inset of Fig.~\ref{fig:lines}(a)

In contrast, the integrals that define the number of photons and the energy of the $\ketspem$ modes converge properly. The convergence of the single-side integrals is already expected from the behavior of the integrands: The photon density $k|\alpha_{sp}(k)|^2$ vanishes as $k\rightarrow 0$ and decreases as $1/k^3$ when $k\rightarrow \infty$, and the energy density $\hbar\cz k^2|\alpha_{sp}(k)|^2$, which also vanishes as $k\rightarrow 0$, behaves as $1/k^2$ when $k\rightarrow \infty$. Actually, when due to the involvement of the two poles, the denominator of $\alpha_{sp}(k)$ contains a polynomial of order two in $k$, the only polynomial order in the numerator that ensures the convergence of both photon number and energy is order zero, as in \Eq{eq:sp2}.

Interestingly, there is one more normalizable mode that can be obtained from \Eq{eq:wspup}. The definition in \Eq{eq:sp2} only uses the term $\frac{-2\ii\text{Im}\left\{q_{sp}\right\}\cz k}{\left(\cz k-\wsp\right)\left(\cz k +\wspstar\right)}$ from \Eq{eq:wspup}, because the other term, $\frac{-2\text{Re}\left\{q_{sp}\wspstar\right\}}{\left(\cz k-\wsp\right)\left(\cz k +\wspstar\right)}$, leads to a divergent normalization integral when taken for each $p$ individually. This, however, changes when taking the sum of all such terms. Let us consider the following mode, to which we assign the index $p=0$:
\begin{equation}
	\label{eq:s0}
	\begin{split}
		&\ket{s0}^\text{em}=\\
		&\frac{1}{\sqrt{n_{s0}}}\intdkmeasure \sum_{p\neq 0}\frac{-2\text{Re}\left\{q_{sp}\wspstar\right\}}{k\left(\cz k-\wsp\right)\left(\cz k +\wspstar\right)}\ketsem.
	\end{split}
\end{equation}
The extra factor of $1/k$ with respect to the integrand in \Eq{eq:sp2} would make $\ket{s0}^\text{em}$ non-normalizable in general, however, the fact that $q_s^2(0)$ must vanish establishes a connection between all the poles and residues:
\begin{equation}
		0=\sum_{p\neq 0}\frac{ 2\text{Re}\left\{q_{sp}\wspstar\right\}}{|\wsp|^2},
\end{equation}
which then ensures that $\ket{s0}^\text{em}$ is normalizable. The energy of $\ket{s0}^\text{em}$ is also finite. The normalization constants and energies of the $0$-th modes, $n_{s0}$, and $\text{H}_{s0}$, respectively, can be obtained by numerical integration for given poles and residues $(q_{sp},\wsp),\ p=1,2,\ldots$:  
\begin{equation}
	\begin{split}
		n_{s0}&=\intdkmeasure \left|\sum_{p\neq 0}\frac{-2\text{Re}\left\{q_{sp}\wspstar\right\}}{k\left(\cz k-\wsp\right)\left(\cz k +\wspstar\right)}\right|^2,\\
		\text{H}_{s0}&=\intdkmeasure \frac{\left(\hbar\cz k\right) }{n_{s0}}\left|\sum_{p\neq 0}\frac{-2\text{Re}\left\{q_{sp}\wspstar\right\}}{k\left(\cz k-\wsp\right)\left(\cz k +\wspstar\right)}\right|^2.
	\end{split}
\end{equation}

Since all the poles in $\ket{s0}^\text{em}$ have negative imaginary part, it is also a polychromatic pulse with a starting time. While arguably its character is somewhat intriguing, there is no obvious reason for {\em not} considering $\ket{s0}^\text{em}$ when formulating theories of emission.

\subsection{The total emission}
In this framework, the total emission $\ket{e}$ is composed of trains of the $\ketspem$ pulses:
\begin{equation}
	\label{eq:esp}
		\ket{e}=\sum_s \sum_{p} \sum_n \exp(\ii\phi_{spn})\op{U}(d_{spn})\ketspem,
\end{equation}
where $\op{U}(\cdot)$ is the time-translation operator, and $\phi_{spn}$ and $d_{spn}$ are the phase and delay of a given pulse.

The statistical properties of $\phi_{spn}$ and $d_{spn}$ will determine the statistics of the emitted light, such as the photon bunching in fully incoherent radiation or the photon anti-bunching in non-classical single photon sources. The way to the set the statistical properties of $\phi_{spn}$ and $d_{spn}$ for obtaining the desired kind of photon statistics is left for future work.

In order to illustrate the formulation of an emission theory with this framework, we will now make very simple assumptions for the statistical properties of $\phi_{spn}$ and $d_{spn}$, and apply the resulting theory to thermal radiation. While the result of this simple example is not fully satisfactory mostly because of the high-frequency tail of the predicted spectra, the next section illustrates general beneficial properties of polychromatic emission theories, shows how randomness is easily incorporated into the formalism, and contains the expression of the Planckian bath in the Hilbert space setting, which is derived in Appendix~\ref{app:phithermal}.

\section{An example for thermal radiation\label{sec:model}}
We now assume that $\phi_{spn}$ are independent random phases with a uniform distribution in the $[-\pi,\pi]$ interval, and that the $d_{spn}$ are independent random times, also independent of the $\phi_{spn}$. With such assumptions, the radiations through the different $\ketspem$ are mutually uncorrelated, as are different photons emitted though a given $\ketspem$. However, each emitted photon has the spatial, polarization, temporal, and frequency coherences of one of the corresponding $\ketspem$ modes. The connection between \Eq{eq:esp} and the photon emission rates for each $\ketspem$ mode is shown in Appendix~\ref{sec:stats}. Essentially, the expected value of the number of emitted photons, $\exval{\braket{e|e}}$, in one second is the sum of the photon emission rates in each mode $\ketspem$, which we denote by $\gammaspem$.

The $\gammaspem$ must ensure that the rate of energy emission out of the $s$-subspace is equal to the rate of energy absorption into such subspace:
\begin{equation}
	\label{eq:econ}
\sum_p \gammaspem\Hsp=\tausabs.
\end{equation}

To fully determine the $\gammaspem$, one needs an extra condition. To such end, we will make another simple assumption and take a Boltzmann distribution, as often done in the context of thermal radiation:
\begin{equation}
	\label{eq:crit}
	\gammaspem=A\exp\left(-\beta\Hsp\right)\impliesduetoeq{eq:econ} A=\frac{\tausabs}{\sum_{\bar{p}} \Hspbar \exp\left(-\beta\Hspbar\right)},
\end{equation}
where $\beta=1/(k_BT)$ is defined as usual with the Boltzmann constant $k_B$. 

Random fields are inherent in thermal processes. One particularly important example is the electromagnetic Planck bath at temperature $T$, which models the illumination experienced by an isolated object in thermal equilibrium with the surrounding photonic bath. The Planckian can be used in more general situations after making the following common assumption, already used by Kirchhoff (see \cite[\S~2]{Baltes1976}): the thermal radiation spectrum of an object at constant temperature $T$ is independent of its environment. Then, the thermal radiation of a hot object heated by any means up to a temperature $T$ above ambient temperature, can be computed by assuming that the object is in thermal equilibrium with a Planckian bath at temperature $T$, and then computing its corresponding radiation.

Random illumination fields can be readily accommodated in the formalism by the substitution:
\begin{equation}
	|f_s(k)|^2\rightarrow\exval{|f_s(k)|^2}.
\end{equation}

For the Planckian, we show in Appendix~\ref{app:phithermal} that:
\begin{equation}
	\exval{|f_s(k)|^2}=\frac{\cz}{2\pi k\left(\exp\left(\frac{\hbar \cz k }{k_B \text{T}}\right)-1\right)},
\end{equation}
with which the $\tausabs$ in \Eq{eq:tausabs} and then the $\gammaspem$ in \Eq{eq:crit} can be computed, thereby fully characterizing the thermal emission from the object predicted with this exemplary theory.

For example, the total rate of emitted photons is
\begin{equation}
	\gammaem=\sum_{sp} \gammaspem =\sum_{sp} \frac{\exp\left(-\beta\Hsp\right)\tausabs}{\sum_{\bar{p}} \Hspbar \exp\left(-\beta\Hspbar\right)}.
\end{equation}

Regarding thermal spectra, the density of the rate of emitted photons per angular wavenumber per second for each $\ketspem$, namely $\gamma_{sp}k|\alpha_{sp}(k)|^2$, may be summed over all the $(s,p)$ to obtain the total spectrum density as a function of $k$:
\begin{equation}
	\label{eq:dkpoly}
	\begin{split}
		&\mathcal{I}_{\text{poly}}(k)=\\
		&\sum_{s,p\neq0} \gamma_{sp}\frac{2\wspre\wspim}{\arctan(\wspre/\wspim)}\frac{k\cz^2}{|\left(\cz k-\wsp\right)\left(\cz k +\wspstar\right)|^2}\\&+ \sum_{s,0}\frac{\gamma_{s0}}{n_{s0}}\frac{1}{k}\left|\sum_{p\neq 0}\frac{-2\text{Re}\left\{q_{sp}\wspstar\right\}}{\left(\cz k-\wsp\right)\left(\cz k +\wspstar\right)} \right|^2.
	\end{split}
\end{equation}
Equation~(\ref{eq:dkpoly}) has some similarities with \cite[Eq.~(10)]{Morino2017}, where the thermal spectrum is also obtained as the superposition of modal emissions. The potential frequency correlations were not discussed in that work. In contrast to ours, the formalism in \cite{Morino2017} is only applicable to spheres and the emission lines are assumed to be Lorentzian.

The angular spectrum at angular wavenumber $\pp$ and direction $\phat$, computed as $k\exval{|\langle \lambda \pp\ket{e}|^2}$, can be readily obtained by adapting \cite[Eq.~(13)]{Mazo2025}, and compared, for example, with what one obtains from the typical use of the directional Kirchhoff law \cite[Eq.~(18)]{Mazo2025}.

We will now compare the density of emission rate per angular wavenumber computed with this polychromatic theory [\Eq{eq:dkpoly}], with its counterpart in the monochromatic theory, whose result can be computed as:
\begin{equation}
	\label{eq:mono}
	\begin{split}
		\mathcal{I}_{\text{mono}}(k)&=\sum_s q_s^2(k)\frac{\cz}{2\pi \left(\exp\left(\frac{\hbar \cz k }{k_B \text{T}}\right)-1\right)}=\\
	&\trace{\matr{{\op{Q}}}(k)}\frac{\cz}{2\pi \left(\exp\left(\frac{\hbar \cz k }{k_B \text{T}}\right)-1\right)},
	\end{split}
\end{equation}
because in the monochromatic theory the densities of absorbed and emitted photons per frequency are identical.

\subsection{A silicon carbide sphere}
As an example, we analyze the thermal emission spectrum of a silicon carbide (SiC) microsphere. The sphere is embedded in air and has a \SI{2}{\micro\meter} radius. In order to use \Eq{eq:dkpoly} and \Eq{eq:crit} we need the poles $\omega_{sp}$ and residues $q_{sp}$ for each multipolar degree (corresponding to the $s$-subspaces). We find these by calculating the S-matrix coefficients  from the Mie coefficients \cite{Wu2012}, and finding their poles and zeros using the AAA algorithm \cite{Hofreither2021} as implemented in the \texttt{baryrat} Python library. In terms of \Eq{eq:csk}, the resonance frequencies $\omega_m$ are the poles, and the residues $c_m$ can be calculated from the poles and zeros using \cite[Eq.~(5)]{Grigoriev2013}. Then, the Q-operator residues $q_{sp}$ are calculated using \Eq{eq:qskapp}.

To do so efficiently, the AAA algorithm needs to be able to calculate the Mie coefficients at complex frequencies, which necessitates the use of an analytical model for the material's permittivity. We took the experimental data for thin-film SiC from Ref.~\cite{Larruquert2011} as a basis for fitting. Considering that SiC is a semiconductor material with a band structure, we approximate the permittivity using the following sum of (two-pole) Lorentzians:
\begin{equation}
	\epsilon(\omega) = \epsilon_\infty + \frac{\omega_{p,\text{IR}}^2}{\omega_{0,\text{IR}}^2 - \omega^2 - i \omega \gamma_\text{IR}} + \sum_n \frac{\omega_{p,n}^2}{\omega_{0,n}^2 - \omega^2 - i \omega \gamma_n} .
\end{equation}
Here, the subscript IR refers to a prominent infrared resonance at 22.4 THz, while the rest of the terms approximate the continuous band structure. In this fit, we have $\epsilon_\infty$ = 1.224. For the IR resonance we have $\omega_{p,\text{IR}}$ = $2 \pi \times$ 36.3 THz, $\omega_{0,\text{IR}}$ = $2 \pi \times$ 22.4 THz, and $\gamma_{\text{IR}}$ = \SI{5.62e13}{\second^{-1}}. We set $\gamma_n$ = \SI{1.13e15}{\second^{-1}} and $\omega_n = n \Delta \omega$ where $\Delta \omega$ = $2 \pi \times$ 120 THz with $n = 1, 2, \dots, 50$. We then use \texttt{minimize} from the \texttt{scipy.optimize} library to fit the oscillator strengths $\omega_{p,n}^2$, minimizing the sum of absolute errors between the experimental data and the fit.

\begin{figure}
	\includegraphics[width=\linewidth]{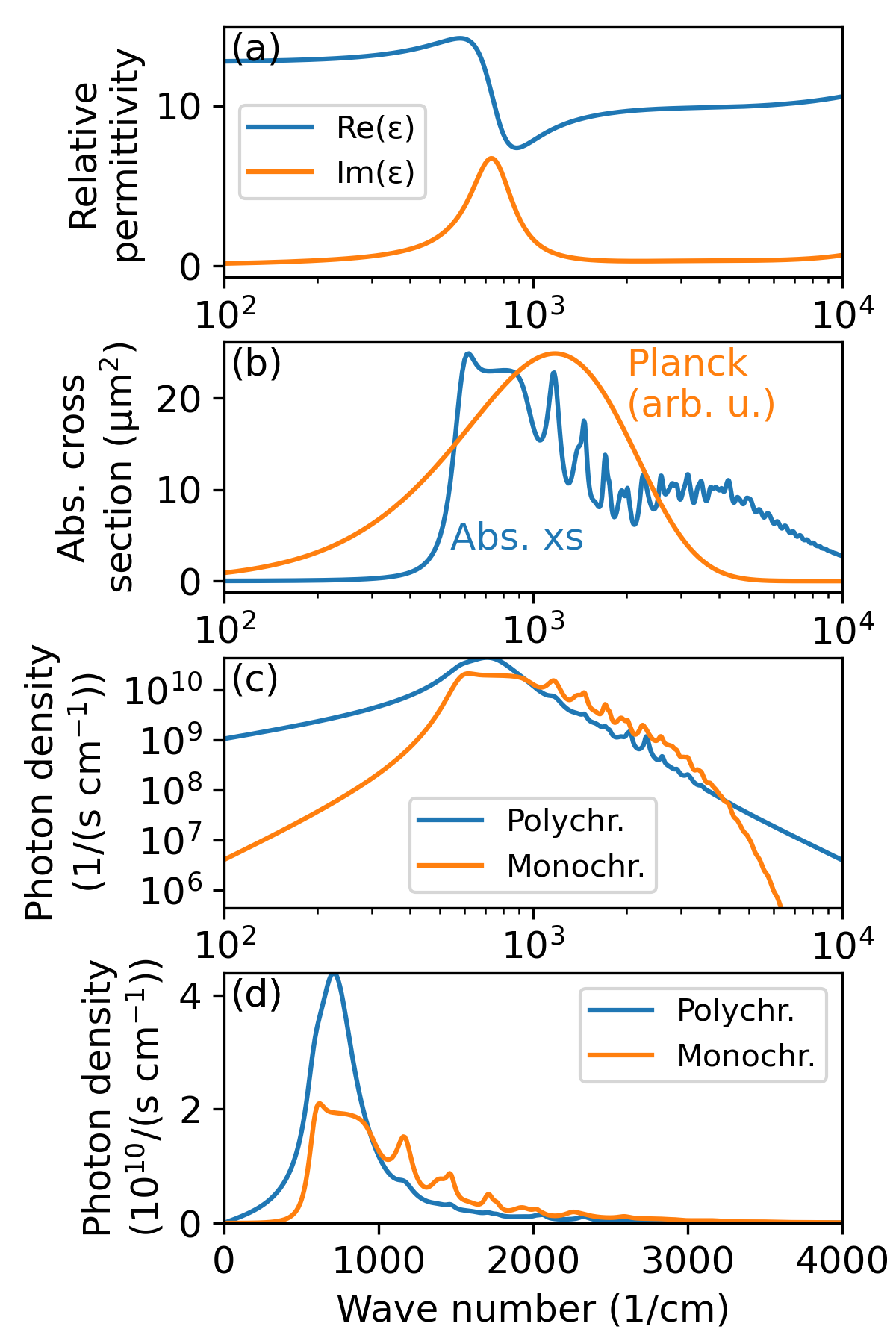}
	\caption{Properties of a \SI{2}{\micro\meter} SiC sphere, as a function of the wavenumber $k/(2\pi)$. Panel (a) shows the real and imaginary parts of the permittivity (blue and orange lines, respectively). Panel (b) shows the absorption cross section (blue line) and the Planckian energy density (orange line) at \SI{600}{\kelvin}. Panel (c) shows the thermal emission spectrum (precisely, the spectral density of photon emission rate) for the poly- and monochromatic theories (blue and orange lines, respectively). Panel (d) shows the data of panel (c) in linear scale and for a reduced range of frequencies.}
	\label{fig:sicsphere1}
\end{figure}

We calculate the resonances up to 6th multipolar order, which is enough to describe the sphere up to about \SI{10000}{\centi\meter^{-1}} frequency. Figure~\ref{fig:sicsphere1}(a) shows the permittivity of SiC in the frequency range of interest, and Fig.~\ref{fig:sicsphere1}(b) shows the absorption cross section along with the Planckian energy density at 600 K temperature. Figures~\ref{fig:sicsphere1}(c,d) show the thermal emission spectrum for both the formulated polychromatic theory and the traditional, monochromatic Kirchhoff approach, in logarithmic and linear scale, respectively. The central region of the spectrum where the absorption cross section overlaps with the Planck spectrum is fairly similar between the two theories. Key differences between the theories include different tail behaviours in both the low- and high-frequency parts. Figure.~\ref{fig:sicsphere1}(c) shows that, while the Planck spectrum and thus the monochromatic-theory emission spectrum fall off more or less exponentially as $k\rightarrow\infty$, this polychromatic theory predicts a $k^{-3}$ tail for the spectrum.
\begin{figure}
	\includegraphics[width=\linewidth]{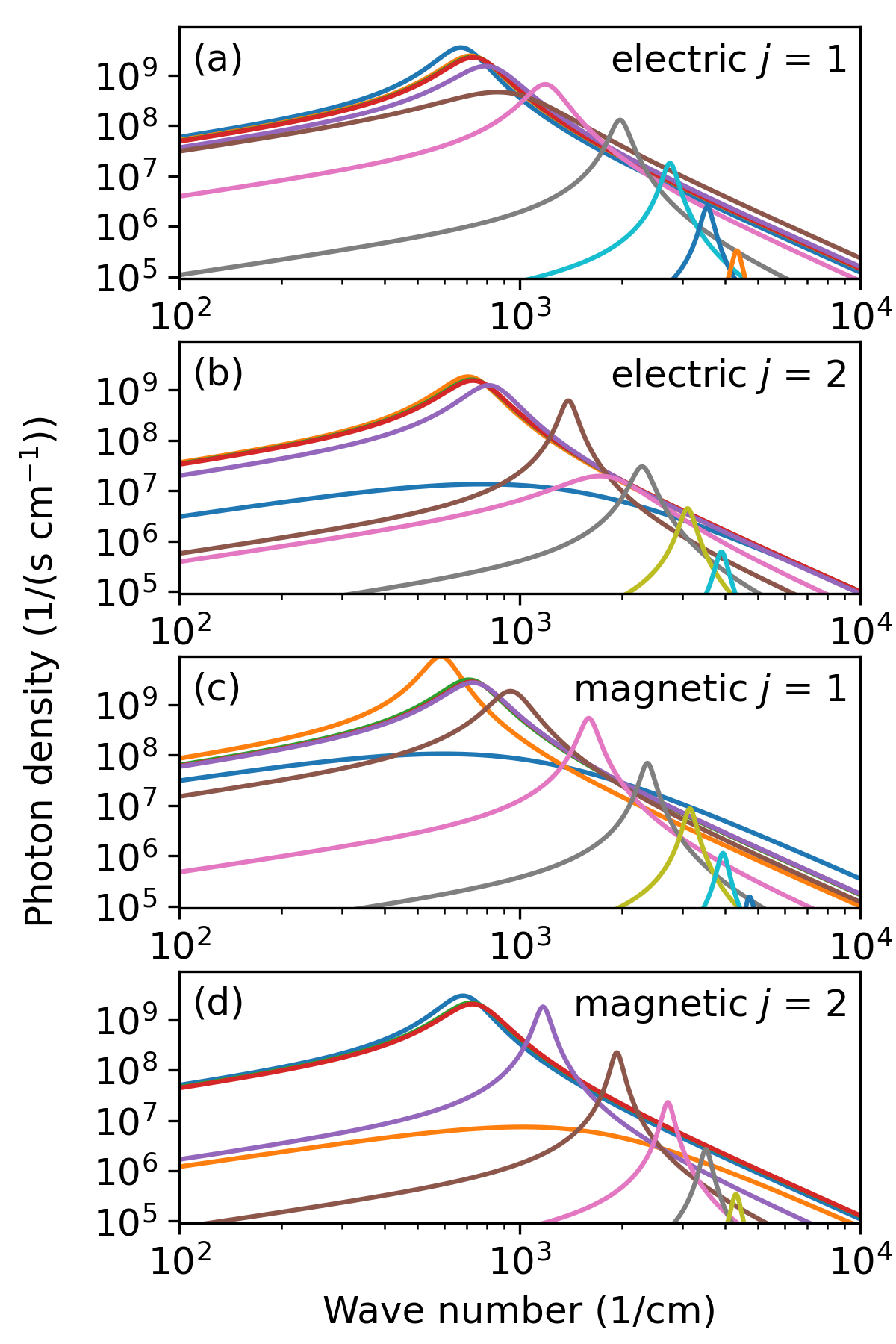}
	\caption{The contribution to the polychromatic-theory emission spectrum [see Fig.~\ref{fig:sicsphere1}(c,d)] by different $q$-resonances. Each resonance's contribution is plotted with a line of a different color. Contributions by the electric dipole and quadrupole are shown in panels (a) and (b), while those of the magnetic dipole and quadrupole are shown in panels (c) and (d), respectively.}
	\label{fig:sicsphere2}
\end{figure}
To elucidate where these differences come from, for the polychromatic theory, we plot the individual contributions to the emission spectrum from each different $q$-resonance. These are shown in Fig.~\ref{fig:sicsphere2}. Each resonance contributes a lineshape like the one in Fig.~\ref{fig:lines}(a), whose peak is close to the real part of the resonance frequency. The Boltzmann factors of \Eq{eq:crit} put more energy into the lower-frequency resonances; we observe that the peak contributions from higher-frequency resonances fall off approximately exponentially with respect to the frequency. The tails of the spectrum are made of the sum of all the individual contributions' tails, all of which have the same $k^{-3}$ dependence. As might be expected for a dielectric sphere, the lowest-frequency resonance has a magnetic dipole character, and this contributes the largest peak in the spectrum [Fig.~\ref{fig:sicsphere2}(c)]. The contribution of the $\ket{s0}^\text{em}$ modes has been computed, but it is so small in this system that it does not appear in the plots.

The polychromatic theory's high-frequency tail is problematic, as it would put substantial energy in the visible and ultraviolet parts of the spectrum even at low temperatures. This runs contrary to observation \cite{Rohlfing1988}. It is a consequence of the spectrum being a sum of relatively wide two-pole Lorentzians, which suggests that some modification of the theory is necessary. One modification is to adapt the statistics of the phases and delays of the emission pulses to reproduce the ``bunching'' characteristics of thermal radiation \cite{Herz2023}.

Importantly, the T-matrices from which one derives the $\op{Q}$ operator and then the $\ketspem$ modes, are typically obtained independently of the temperature. This approximation is very often used in computations of thermal radiation, where material parameters measured at room temperature are used to describe the systems at different temperatures. Similarly, molecular T-matrices for the ground state are used \cite{Mazo2025}, ignoring for example the population of vibrational states at room temperature. These approximations should be addressed for improving the predictions of any emission theory.

\section{Properties of polychromatic emission theories\label{sec:properties}}
\begin{figure}
	\includegraphics[width=\linewidth]{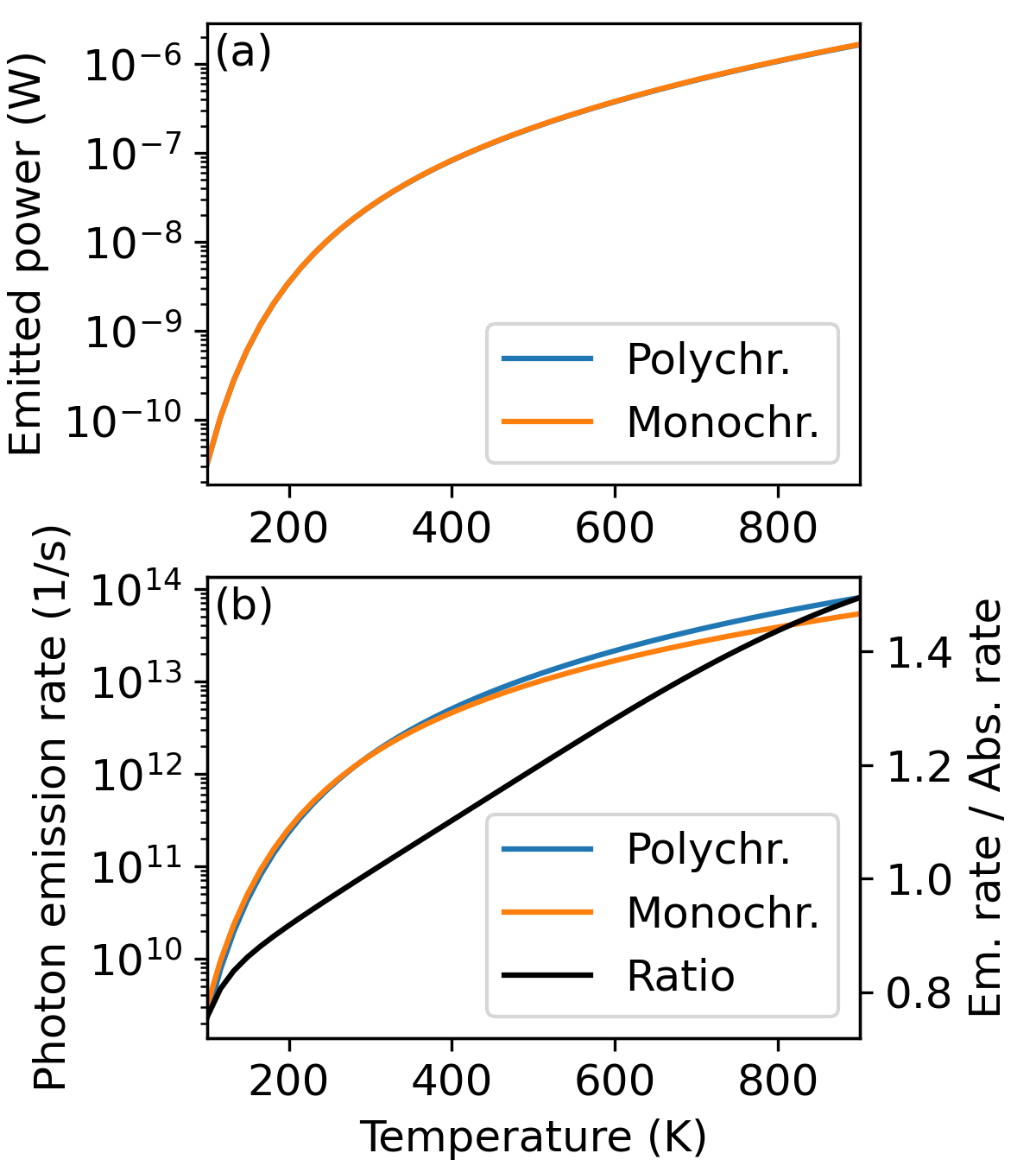}
	\caption{Temperature-dependent emission by the \SI{2}{\micro\meter} SiC sphere. Panel (a) shows the emitted power as a function of temperature; both the polychromatic theory (blue line) and monochromatic theory (orange line) give the same result. Panel (b) shows the photon emission rate as a function of temperature, with the polychromatic and monochromatic results (blue and orange lines, respectively) on the left y-axis. The black curve shows, for the polychromatic theory, the ratio between emitted and absorbed photons. For the monochromatic theory, this ratio is always unity.}
	\label{fig:sicsphere3}
\end{figure}

Emission theories built with the polychromatic $\ketspem$ modes have the ability to produce significant radiation in frequencies that are much suppressed in or even absent from the illumination, as seen in Fig.~\ref{fig:sicsphere1}. The differences between the absorbed an emitted spectra, together with energy conservation, imply that the ratio between the rates of emitted and absorbed photons is in general different than one. This is illustrated in Fig.~\ref{fig:sicsphere3} for the SiC sphere. Both theories give the same result when looking at the emitted power [Fig.~\ref{fig:sicsphere3}(a)], because in both cases the absorbed power from the Planckian bath is conserved. In terms of the photon emission rate, however, the theories give different predictions [Fig.~\ref{fig:sicsphere3}(b)]. For the polychromatic theory, the ratio between the number of emitted and absorbed photons [black line in Fig.~\ref{fig:sicsphere3}(b)] is given by:
\begin{equation}
	\frac{\gammaem}{\gammaabs}=\frac{\sum_{sp}\gammaspem}{\gammaabs}=\frac{\sum_{sp}\gammaspem}{\sum_s \intdkmeasure q_s^2(k)\exval{|f_s(k)|^2}},
\end{equation}
and tends to be less than unity for low temperatures and larger than unity for high temperatures. This is explained by the fact that at a low temperature the Boltzmann factors of \Eq{eq:crit} assign almost all the absorbed energy to the lowest-frequency resonance. If this resonance lies above the peak of the low-temperature Planckian spectrum, we expect that many low-energy photons are absorbed but the energy is then re-emitted at the relatively higher photon energy of the resonance. Vice versa, at high temperatures the Boltzmann factors are still largest for the lower-frequency resonances, and the high-energy photons absorbed from the Planckian bath are converted into a larger number of lower-energy photons in the emission. 
This is in contrast with monochromatic theories where the energy absorbed by the object at each frequency is returned to the environment into the same frequency. Since in such case the density of emitted and absorbed photons per unit frequency is identical, the ratio of their respective integrals, that is $\frac{\gammaem}{\gammaabs}$, is equal to one. 

In many realistic situations, the rates of absorbed and emitted photons are different. For example in the case of a nanoparticle heated up to a high temperature by a laser in the visible, there are many more photons emitted in the infrared than absorbed in the visible. With the polychromatic approach, we can treat the situation as follows. The energy absorbed in each subspace $s$, namely the $\tau_s^\text{abs}$ in \Eq{eq:tausabs}, is computed for any given illumination $\ket{f}$ using the quantities $|f_s(k)|^2=|\brasabs f\rangle|^2$. Then, instead of the exponential factors $\exp\left(-\beta\Hsp\right)$ in \Eq{eq:crit}, one may, for example, substitute them by the overlap between the illumination and a polychromatic incoming mode: $|{}^\text{abs}\bra{ps}f\rangle|^2$. Such a polychromatic mode is obtained by substituting $\ketsem$ for $\ketsabs$ in \Eq{eq:sp2}.

The ability to handle general illuminations in this direct way, producing emissions with frequencies that are much suppressed in or even absent from the illumination, and with different rates of emission and absorption, make the polychromatic framework suitable for describing different kinds of emissions, such as for example luminescence, in the Hilbert space.

\section{Conclusion and outlook\label{sec:outlook}}
We have outlined a framework for formulating polychromatic theories of emission in the electromagnetic Hilbert space. The emission is modeled as trains of polychromatic pulses. Such pulses are derived using the natural resonance frequencies of the object, and have conceptual advantages over the typical Lorentzian lineshapes.

Three research topics will be addressed in the future. One is studying how to set the statistical properties of the emission, such as ``bunching'' and other options, through the statistics of the delays and phases of the emitted pulses. Another is obtaining absorption operators of objects at specified temperatures. For molecules, this implies the inclusion of vibronic couplings, which modify the absorption spectrum. For materials, this implies the consideration of thermally excited atoms in the derivation of the material parameters. The third topic is the formulation of a polychromatic theory of luminescence in the Hilbert space.

\begin{acknowledgments}
We thank Juan Diego Mazo-V\'asquez and Jan David Fischbach for useful discussions. I.F-C acknowledges funding by the Helmholtz Association via the Helmholtz program ``Materials Systems Engineering'' (MSE). M.V. acknowledges funding by the Deutsche Forschungsgemeinschaft (DFG, German Research Foundation) -- Project-ID 258734477 -- SFB 1173. M.N. acknowledges support by the KIT through the ``Virtual Materials Design'' (VIRTMAT) project. 
\end{acknowledgments}

\newpage
\appendix

\section{Mathematical setting\label{app:conventions}}
The set of solutions of Maxwell equations, together with the electromagnetic scalar product \cite{Gross1964}, form a Hilbert space, which we denote by $\mathbb{M}$. The scalar product between two members of $\mathbb{M}$ can be computed as:
\begin{equation}
	\label{eq:lmsp}
	\langle f|g\rangle = \sum_{\lambda=\pm1} \int_{\mathbb{R}^3-\zerovec} \frac{\text{d}^3 \pp}{k} \, \ffestar{\lambda}g_\lambda (\pp),
\end{equation}
where $\pp$ is the wavevector, $\lambda$ the polarization handedness, and $\ff{\lambda}$ and $\mathrm{g}_\lambda (\pp)$ are the coefficient functions of the plane wave expansions of $\ket{f}$ and $\ket{g}$, respectively. The scalar product can also be computed as
\begin{equation}
	\label{eq:amsp}
	\langle f|g\rangle = \sum_{\lambda=\pm 1} \int_{>0}^{\infty} \text{d}k\, k \sum_{j=1}^{\infty} \sum_{m=-j}^j \, \FFestar{\lambda} {g}_{jm\lambda}(k),
\end{equation}
where the $\FFe{\lambda}$ and ${g}_{jm\lambda}(k)$ are the coefficient functions of the expansions in spherical waves, also known as multipolar fields. The form of the expressions (\ref{eq:lmsp}) and (\ref{eq:amsp}) is achieved using the conventions in \cite{Vavilin2023}, which we include here for reference.

The electric field of a particular solution $\ket{f}$ is expanded into plane waves of well-defined helicity $\ket{\pp \lambda}$ as:
\begin{equation}
	\label{eq:xpans}
	\Ert=\sum_{\lambda=\pm1 }\int_{\mathbb{R}^3-\zerovec} \frac{\text{d}^3 \pp}{k} \, \ffe{\lambda} \, \ket{\pp\lambda},
\end{equation}
and the plane waves are defined as:
\begin{equation}
	\begin{split}
		&\ket{\pp \lambda}\equiv\\
		&\sqrt{\frac{\cz\hbar}{ \epsz}}\, \frac{1}{\sqrt{2}} \frac{1}{\sqrt{(2\pi)^3}}\, k \, \mathbf{\hat{e}}_\lambda({\phat}) \exp(- \ii k\cz t ) \exp(i \pp \cdot \rr),
	\end{split}
\end{equation}
where $\cz$ is the speed of light in vacuum, $\hbar$ the reduced Planck constant, and $\epsz$ the permittivity in vacuum. We highlight the factor of $k=|\pp|$ in the definition of the plane waves, which ensures that they transform unitarily under Lorentz transformations, and the factor of $1/k$ in \Eq{eq:xpans}, which renders the volume measure $\frac{\text{d}^3 \pp}{k}$ invariant under transformations in the Poincar\'e group.

The expansion in multipoles of well-defined helicity reads:
\begin{equation}
	\label{eq:ypans}
	\begin{split}
		&{\Ert}^{\text{reg/in/out}} \equiv\\
		&\int_{>0}^\infty \text{d}k \, k \, \sum_{\lambda=\pm 1} \sum_{j=1}^{\infty} \sum_{m=-j}^j \, \FFe{\lambda} \, \ket{k j m \lambda}^{\text{reg/in/out}},
	\end{split}
\end{equation}
where $j=1,2,...$ is the multipolar degree with $j=1$ corresponding to dipoles, $j=2$ to quadrupoles, and so on, and $m=-j,-j+1,\ldots, j$ is the component of the angular momentum in the $z$ direction.

The regular, incoming, and outgoing multipoles $\ket{k j m \lambda}^{\text{reg/in/out}}$ are defined as:
\begin{widetext}
	\begin{equation}
		\begin{split}
		\label{eq:mpdef}
			&\ket{kjm\lambda}^{\text{reg}}\equiv\mathbf{S}^\text{reg}_{jm\lambda}(k,\rr,t)=\\
			&- \sqrt{\frac{\cz\hbar}{\epsilon_0}} \frac{1}{\sqrt{2\pi}} \, k \, \ii^j  \times\Big(  \exp(-\ii k \cz t)\, \mathbf{N}^{\text{reg}}_{jm}(k|\rr|, \hat{\rr}) + \lambda \,\exp(-\ii k\cz t) \,  \mathbf{M}^{\text{reg}}_{jm}(k|\rr|, \rhat ) \Big),\\
			&\ket{kjm\lambda}^{\text{in/out}}\equiv\mathbf{S}^\text{in/out}_{jm\lambda}(k,\rr,t)=\\
			&- \frac{1}{2} \sqrt{\frac{\cz\hbar}{\epsilon_0}} \frac{1}{\sqrt{2\pi}} \, k \, \ii^j\times \Big(  \exp(-\ii k \cz t)\, \mathbf{N}^{\inout}_{jm}(k|\rr|, \hat{\rr}) + \lambda \,\exp(-\ii k\cz t) \,  \mathbf{M}^{\inout}_{jm}(k|\rr|, \rhat ) \Big),
		\end{split}
	\end{equation}
\end{widetext}
where the $\mathbf{M}$ and $\mathbf{N}$ have the usual definitions (see e.g. \cite[Eqs.~(50,51)]{Vavilin2023}).

The expansions in \Eq{eq:xpans} and \Eq{eq:ypans}, together with the requirements
\begin{equation}
	\bra{\barlambda\mbar\jbar p}f\rangle=f_{\jbar\mbar\barlambda}(p),\text{ and } \bra{\barlambda\qq}f\rangle=f_{\barlambda}(\qq),
\end{equation}
impose the conditions
\begin{equation}
	\label{eq:normswpw}
	\begin{split}
		\langle \barlambda \mbar \jbar p|k j m \lambda\rangle&=\delta_{\jbar j}\delta_{\mbar m}\delta_{\barlambda\lambda}\frac{\delta(p-k)}{k},\text{ and}\\
		\langle \barlambda \qq|\pp \lambda\rangle&=\delta_{\lambda\barlambda}\delta(\pp-\qq)k.
	\end{split}
\end{equation}

One advantage of using the electromagnetic Hilbert space for the study of light-matter interaction is that the amount of fundamental quantities contained in a given field is easily formulated: $\sandwich{f}{\Gamma}{f}$, where $\Gamma$ is the self-adjoint operator representing the fundamental quantity, for example energy or momentum. In the multipolar basis, the energy of the field would be computed as:

\begin{equation}
	\label{eq:last}
	\sandwich{f}{\op{H}}{f}=\sum_{jm\lambda}\intdkmeasure \left(\hbar\cz k\right)|\FFe{\lambda}|^2.
\end{equation}

\section{Expansion of $q_s^2(k)$ into partial fractions\label{app:qfroms}}
The following expression for the scattering operator $S$ can be obtained with steps similar to those leading to \Eq{eq:qkknew}:
\begin{equation}
	\label{eq:skknew}
		\op{S}=\sum_{jm\lambda}\sum_{\jbar\mbar\barlambda}\intdkmeasure \ \op{S}^{jm\lambda}_{\jbar\mbar\barlambda} (k) |k j m \lambda\rangle\langle \barlambda\mbar\jbar k|.
\end{equation}
Then, the SVD at each $k$
\begin{equation}
	\label{eq:svdsknew}
	\matr{{\op{S}}}(k)=\sum_s c_s(k) \vec{a}_s(k)\vec{b}_s^\dagger(k),
\end{equation}
and the relationship $\matr{{\op{Q}}}(k)=\Id-\matr{{\op{S}}}^\dagger(k)\matr{{\op{S}}}(k)$ imply that:
\begin{equation}
	 q_s^2(k)=1-|c_s(k)|^2.
\end{equation}

We will now use an expansion into partial fractions that was obtained in \cite{Grigoriev2013} for the S-matrix coefficients of spherical scatterers, namely \cite[Eq.~(4)]{Grigoriev2013}:
\begin{equation}
	m_s(k)=a_s\exp(\ii b_s k)\left(1+\sum_m \frac{c_m}{\cz k-\omega_m}\right).
\end{equation}
Since $m_s(k)$ and $c_s(k)$ can only differ by a complex phase, we have that:
\begin{equation}
	\label{eq:csk}
	|c_s(k)|^2=\left|a_s\exp(\ii b_s k)\left(1+\sum_m \frac{c_m}{\cz k-\omega_m}\right)\right|^2.
\end{equation}
While we are dealing with objects of general shape, the key property used in \cite{Grigoriev2013} is that the scattering channels are decoupled. Spherical non-chiral scatterers ensure such decoupling for the electric and magnetic multipolar fields. In our case, we achieve such decoupling through the $\vec{a}_s(k)\vec{b}_s^\dagger(k)$ projectors from \Eq{eq:svdsknew}. 

It is readily seen from \cite[Eq.~(3)]{Grigoriev2013} that $b_s$ is a real number, because the poles come in pairs $(\omega_m,-\omega_m^*)$. Causality forces the imaginary part of all the poles to be negative, and we can assume that the $\omega_m(-\omega_m^*)$ have positive(negative) real part. Moreover, \cite[Eq.~(5)]{Grigoriev2013} implies that the residue of the $-\omega_m^*$ pole is equal to minus the complex conjugate of the residue of the $\omega_m$ pole.

We can then write:
\begin{equation}
	\label{eq:qskapp0}
	\begin{split}
		&q_s^2(k)=1-|c_s(k)|^2=1-|a_s|^2-\\
		&|a_s|^2\left[\sum_m \left(\frac{c_m}{\cz k-\omega_m}+ \frac{c_m^*}{\cz k-\omega_m^*}\right)+\left|\sum_n\frac{c_n}{\cz k-\omega_n}\right|^2\right],
	\end{split}
\end{equation}
which can be shown to be:
\begin{equation}
	\label{eq:qskapp}
	\begin{split}
		&q_s^2(k)=1-|c_s(k)|^2=1-|a_s|^2-|a_s|^2\times\\
		&\sum_m \left(\frac{q_m}{\cz k-\omega_m}-\frac{q_m^*}{\cz k+\omega_m^*}+\frac{q_m^*}{\cz k-\omega_m^*}-\frac{q_m}{\cz k+\omega_m}\right),\\
		&\text{ with } q_m=c_m+\sum_n \frac{c_mc_n^*}{\omega_m-\omega_n^*}-\frac{c_mc_n}{\omega_m+\omega_n}.
	\end{split}
\end{equation}
	In Eqs.~(\ref{eq:qskapp0},\ref{eq:qskapp}), and from now on, the sums over the pole indexes, such as the sums over $n$ and $m$ above, run only over the $\wsp$, with positive real part and negative imaginary part.

To finish, we will use that the absorption tends to zero as $k$ grows to infinity: $\lim_{k\rightarrow\infty}q_s^2(k)=0$, which implies, from \Eq{eq:qskapp}, that $|a_s|^2=1$. Therefore: 
\begin{equation}
\begin{split}
	&-q_s^2(k)=\\
		&\sum_m \left(\frac{q_m}{\cz k-\omega_m}-\frac{q_m^*}{\cz k+\omega_m^*}+\frac{q_m^*}{\cz k-\omega_m^*}-\frac{q_m}{\cz k+\omega_m}\right).
\end{split}
\end{equation}

\section{The $\ketspem$ modes\label{app:spn}}
\subsection{Normalization}
Let us start by using all the terms in \Eq{eq:sp2} with poles in the lower part of the complex plane:
\begin{equation}
	\label{eq:sp1}
	\begin{split}
		&\ketspem=\\
		&\frac{1}{\sqrt{n_{sp}}}\intdkmeasure \frac{1}{k}\left[\frac{-2\ii\text{Im}\left\{q_{sp}\right\}\cz k-2\text{Re}\left\{q_{sp}\wspstar\right\}}{\left(\cz k-\wsp\right)\left(\cz k +\wspstar\right)}\right] \ketsem
	\end{split}
\end{equation}
, where $n_{sp}$ is a constant which needs to ensure that $\braspem sp\rangle^{\text{em}}=1$, that is, we are after single-photon modes. The factor $\frac{1}{k}$ outside the square brackets ensures the correct units of meters in the coefficients between the integration measure $\int \text{d}kk$ and the $\ket{ks}$. 

The normalization constant $n_{sp}$ in \Eq{eq:sp1} can then be written as:
\begin{equation}
	\label{eq:nsp}
		n_{sp}=\int_{>0}^\infty {\frac{\mathrm{d}k}{k}} \left|\frac{-2\ii\Im{q_{sp}}\cz k-2\text{Re}\left\{q_{sp}\wspstar\right\}}{\left(\cz k-\wsp\right)\left(\cz k +\wspstar\right)}\right|^2.
\end{equation}

It is readily seen that the integral diverges due to the behavior of the integrand as $k\rightarrow 0$, which is caused by the term  $-2\text{Re}\left\{q_{sp}\wspstar\right\}$. We therefore drop such term, and obtain the integral:
\begin{equation}
	\label{eq:nsp2}
	\begin{split}
	n_{sp}&=|2\ii\Im{q_{sp}}|^2\times\\
	&\int_{>0}^\infty \ \mathrm{d}(\cz k) \frac{\cz k}{\left|\left(\cz k-\wsp\right)\left(\cz k +\wspstar\right)\right|^2},
	\end{split}
\end{equation}
which can be solved analytically, resulting in: 
\begin{equation}
	\label{eq:nsp3}
	\begin{split}
		n_{sp}&=|2\ii\Im{q_{sp}}|^2\frac{\arctan(\wspre/\wspim)}{2\wspre\wspim}\text{, and hence} \\
		\frac{1}{\sqrt{n_{sp}}}&=\sqrt{\frac{2\wspre\wspim}{\arctan(\wspre/\wspim)}}\frac{1}{|2\ii\Im{q_{sp}\exp(\ii\phi_{sp})}|},
	\end{split}
\end{equation}
where $\wspre$ and $\wspim$ are the real and imaginary parts of $\wsp$, respectively.

Combining the last line of \Eq{eq:nsp3} with \Eq{eq:sp1} without the excluded divergent term, one obtains:
\begin{equation}
	\label{eq:sp2app}
	\begin{split}
		\ketspem&=-\text{sign}\left[\Im{q_{sp}}\right]\sqrt{\frac{2\wspre\wspim}{\arctan(\wspre/\wspim)}}\times\\
		&\intdkmeasure \frac{\ii\cz}{\left(\cz k-\wsp\right)\left(\cz k +\wspstar\right)}\ketsem.
	\end{split}
\end{equation}
Finally, we set the global phase so that the two-sided Fourier transform produces a real-valued waveform [\Eq{eq:time}], and obtain the definition of $\ketspem$ in \Eq{eq:sp2} in the text, which does not depend on $q_{sp}$ at all. In hinsight, this is not surprising because the modes are normalized.

\subsection{Energy}
The energy of a given $\ketspem$ is readily computed using the action of the energy operator $\op{H}$ on the monochromatic $\ket{ks}$: $\op{H}\ket{ks}=\hbar\cz k\ket{ks}$. Then:
\begin{equation}
	\begin{split}
		&\braspem\op{H}\ketspem=\\
		&\intdkmeasure (\hbar \cz k)\left|\frac{\sqrt{\frac{2\wspre\wspim}{\arctan(\wspre/\wspim)}}\ii\cz}{\left(\cz k-\wsp\right)\left(\cz k +\wspstar\right)}\right|^2,
	\end{split}
\end{equation}
which, we manipulate into
\begin{equation}
	\begin{split}
		&\braspem\op{H}\ketspem=\frac{2\wspre\wspim\hbar}{\arctan(\wspre/\wspim)}\times\\
		&\int_{>0}^\infty {\mathrm{d}(\cz k)} \frac{(\cz k)^2}{\left|\left(\cz k-\wsp\right)\left(\cz k +\wspstar\right)\right|^2}\ ,
	\end{split}
\end{equation}
and, after solving the integral analytically, obtain:
\begin{equation}
	\braspem\op{H}\ketspem=\hbar \wspre\frac{-\pi/2}{\arctan(\wspre/\wspim)}.
\end{equation}

\section{Photon emission rates\label{sec:stats}}
We here derive the connection between \Eq{eq:esp} and the photon emission rates for each $\ketspem$ mode. We start from \Eq{eq:esp}
\begin{equation}
		\ket{e}=\sum_s \sum_{p} \sum_n \exp(\ii\alpha_{spn})\op{U}(d_{spn})\ketspem,
\end{equation}
and write down the expected value of the total number of photons in $\ket{e}$:
\begin{equation}
	\begin{split}
	&\exval{\braket{e|e}}=\sum_{s\bar{s}} \sum_{p\bar{p}} \sum_{n\bar{n}}\\
		&	\exval{\exp(\ii(\alpha_{spn}-\alpha_{\bar{s}\bar{p}\bar{n}}))\op{U}^\dagger(d_{\bar{s}\bar{p}\bar{n}})\op{U}(d_{spn})}\braket{\bar{p}\bar{s}\ketspem}=\\
		&	\exval{\exp(\ii(\alpha_{spn}-\alpha_{\bar{s}\bar{p}\bar{n}}))}\exval{\op{U}^\dagger(d_{\bar{s}\bar{p}\bar{n}})\op{U}(d_{spn})}\braket{\bar{p}\bar{s}\ketspem}=\\
		& \sum_{sp}\sum_n \braspem sp\rangle^{\text{em}},
	\end{split}
\end{equation}
where the second equality follows from assuming that the phases are independent from the delays, and the third from $\exval{\exp(\ii(\alpha_{spn}-\alpha_{\bar{s}\bar{p}\bar{n}}))}=\delta_{s\bar{s}}\delta_{p\bar{p}}\delta_{n\bar{n}}$, and the unitary character of $\op{U}(d)$. 

To finish, we choose a finite time period $D$ so that a finite number of photons are emitted through each $\ketspem$ mode during such period. Such number of photons, $N_{sp}$, is also a random variable. We then have that:
\begin{equation}
	\exval{\braket{e|e}}_D=\sum_{sp}\exval{\sum_{n_s=1}^{N^D_{sp}}1} = \sum_{sp}\exval{N^D_{sp}}.
\end{equation}

If we set $D=1$ second, we can identify $\exval{N^{D=1}_{sp}}$ with the photon emission rate through the $\ketspem$ mode, which we denote by $\gammaspem$ in the text.

\section{The value of $\exval{|f_s(k)|^2}$ for a Planckian thermal bath\label{app:phithermal}}
We start with a plane-wave expansion of the thermal bath:
\begin{equation}
	\label{eq:respa}
	\ketbathe =\sum_{\lambda=\pm 1}\intdpinv \psi_\lambda(\pp)|\pp \lambda\rangle,
\end{equation}
where, in the Planckian case, we assume that the radiation is uncorrelated in frequency, direction, and polarization, and also equal for both polarizations:
\begin{equation}
	\label{eq:corrbathapp}
	\exval{\psi_\lambda(\pp)\psi_{\barlambda}^*(\qq)}=\delta_{\lambda\barlambda}\delta(\pp-\qq)k^3\exval{|\psi_+(k)|^2}.
\end{equation}

To compute $\exval{|\psi_{+}(k)|^2}$, we will assume that an achiral sphere is immersed in the Planckian bath and obtain two different expressions for the total power absorbed by the sphere. Once using the standard connection with the absorption cross-section of the sphere, and once in our setting. The expression for $\exval{|\psi_{+}(k)|^2}$ follows from equating both expressions. Since $\exval{|\psi_{+}(k)|^2}$ is independent of the object, we are free to select any object for this derivation. The high symmetry of an achiral sphere simplifies the work.

The energy density of all the plane waves of the Planckian bath sums up to:
\begin{equation}
	u_\text{Planck}(k) = \frac{\hbar \cz k^3}{\pi^2} \frac{1}{\exp\left(\frac{\hbar \cz k }{k_B \text{T}} k\right) - 1}
\end{equation}
. The wavenumber-dependent intensity incident on the sphere from the Planck bath is then $I(k) = u_\text{Planck}(k) \cz$, and the power absorbed from the Planckian bath in some differential spectral width $\text{d}k$ is then:
\begin{align}
	& \frac{P}{\text{d}k} = u_\text{Planck}(k) \cz \sigma_\text{abs}(k)\nonumber \\
	&           = \frac{\hbar \cz^2 k^3}{\pi^2} \frac{1}{\exp\left(\frac{\hbar \cz k }{k_B \text{T}} k\right) - 1} \sigma_{\text{abs}}(k), \label{debug3}
\end{align}
where $\sigma_{\text{abs}}(k)$ is the absorption cross-section of the sphere:
\begin{equation}
	\label{eq:acss}
	\begin{split}
		&\sigma_\text{abs}(k) =\\
		&\frac{2 \pi}{k^2} \sum_j (2 j + 1) \left(-\text{Re}[a_j(k) + b_j(k)] - |a_j(k)|^2 - |b_j(k)|^2\right),
	\end{split}
\end{equation}
where $a_j(k)$ and $b_j(k)$ are the electric and magnetic Mie coefficients, respectively.

We can obtain the total absorbed power by integrating over the expression in \Eq{debug3} with $\int \text{d}k$: 
\begin{align}
	&	P =\int_{>0}^\infty \text{d}k \frac{\hbar \cz^2 k^3}{\pi^2} \frac{1}{\exp(\frac{\hbar \cz k }{k_B \text{T}}) - 1} \sigma_{\text{abs}}(k).
\label{eq:nop}
\end{align}

We will now obtain a different expression for the same quantity, but involving our target $|\psi_+(k)|^2$. We need to compare the absorbed energy per second in \Eq{eq:nop} with an expression for the energy absorbed by the object in our setting, because we have assumed a reference period of one second, with which energy absorption actually represents energy absorption rate. The sought after expression for energy absorption can be written from \Eq{eq:fQf} following \cite[Appendix~B]{Mazo2025}, by assuming that the $\op{T}(\op{Q})$ operator of the sphere does not mix frequencies, so that $\op{Q}\op{H}=\op{H}\op{Q}$. The combination of Eq.~(B1) with Eq.~(B6) in that reference is:
\begin{equation}
	\begin{split}
	&\exval{\langle \Psi_{\text{bath}}|\op{Q}\ketbathe}=\\
	&\sum_s \intdkmeasure q_s^2(k)\sum_\lambda\exval{|\psi_\lambda(k)|^2}\sum_{jm}|\vec{s}_{jm\lambda}(k)|^2\,
	\end{split}
\end{equation}
which represents the photon absorption, and from which it can readily be seen that the energy absorption can be written as:
\begin{equation}
	\begin{split}
	&\exval{\langle \Psi_{\text{bath}}|\op{Q}\op{H}\ketbathe}=\\
		&\sum_s \intdkmeasure (\hbar \cz k) q_s^2(k)\sum_\lambda\exval{|\psi_\lambda(k)|^2}\sum_{jm}|\vec{s}_{jm\lambda}(k)|^2.
	\end{split}
\end{equation}
Since we have assumed that $\exval{|\psi_\lambda(k)|^2}=\exval{|\psi_+(k)|^2}$, it is possible to continue as follows:
\begin{equation}
	\begin{split}
	&\exval{\langle \Psi_{\text{bath}}|\op{Q}\op{H}\ketbathe}=\\
		&\sum_s \intdkmeasure (\hbar \cz k) q_s^2(k)\exval{|\psi_+(k)|^2}\sum_{jm\lambda}|\vec{s}_{jm\lambda}(k)|^2=\\
		&\intdkmeasure (\hbar \cz k) \exval{|\psi_+(k)|^2}\sum_s q_s^2(k)=\\
		&\intdkmeasure (\hbar \cz k) \exval{|\psi_+(k)|^2}\frac{2k^2\sigma_{\text{abs}}(k)}{\pi},
	\end{split}
\end{equation}
where the last equation follows from \cite[Eq.~(B11)]{Mazo2025}. We can now equate the integrand with the one in \Eq{eq:nop} to obtain:
\begin{equation}
	\label{eq:psisqapp}
	\exval{|\psi_+(k)|^2} = \frac{\cz}{2\pi k\left(\exp\left(\frac{\hbar \cz k }{k_B \text{T}}\right)-1\right)}.
\end{equation}
\begin{figure}[ht!]
	\includegraphics[width=0.5\textwidth]{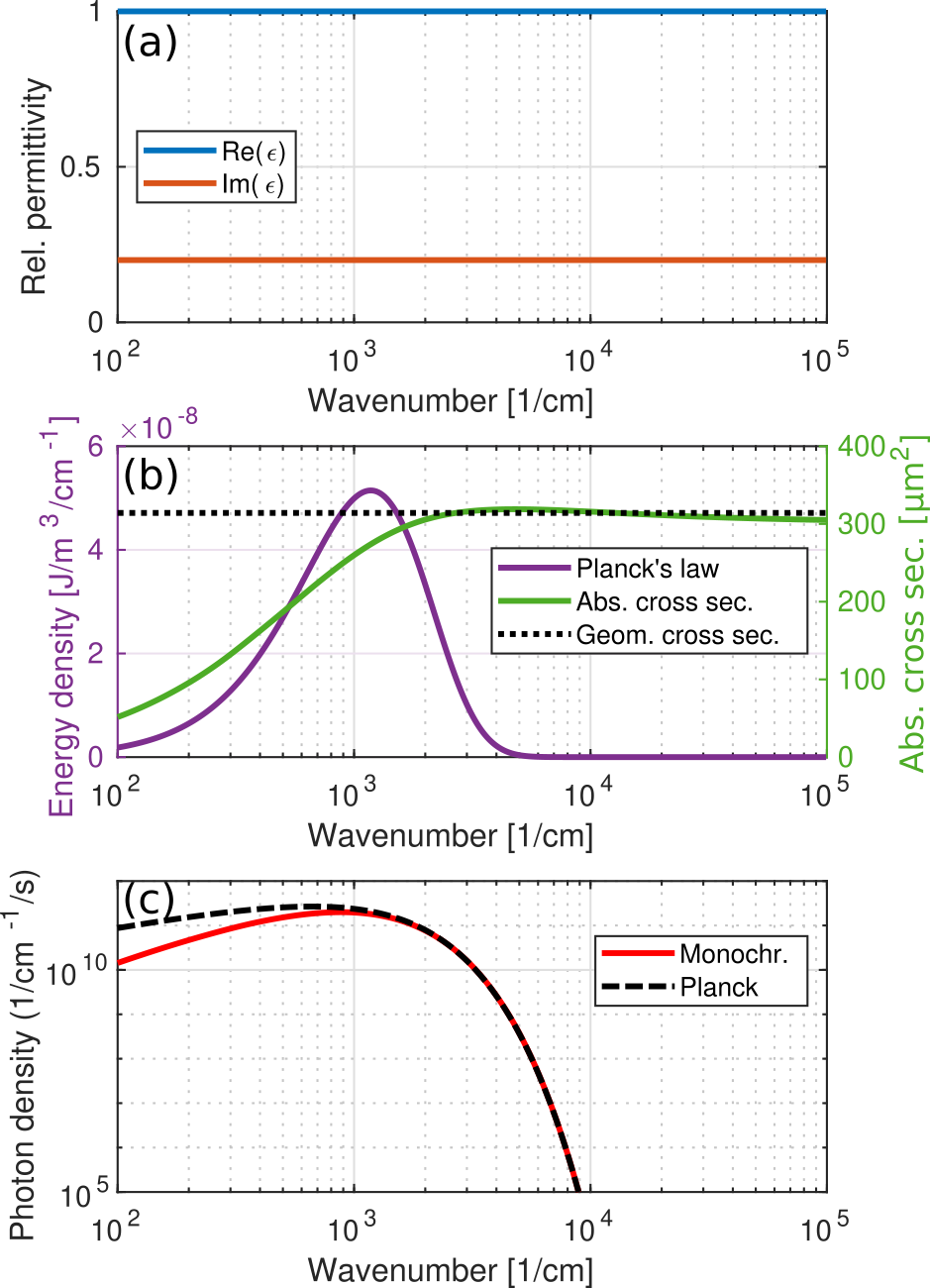}
	\caption{Thermal spectra of a nearly black body. Panel (a) shows the (constant) real and imaginary parts of the permittivity as a function of wavenumber. Panel (b) shows the Planck distribution at 600 K as well as the absorption cross section, which closely follows the geometrical cross section at wavenumbers above \SI{1000}{\centi\meter^{-1}}. Panel (c) shows the thermal radiation spectra calculated the polychromatic and monochromatic theories, along with the Planck spectrum, with the monochromatic result following the Planck spectrum above \SI{1000}{\centi\meter^{-1}}.}
	\label{fig:blackbody}
\end{figure}
One way to verify \Eq{eq:psisqapp} is to use it in the monochromatic theory for computing the thermal radiation spectrum of an object that behaves as a black body in some limit. In such a limit, the spectrum should approach the Planck spectrum. To verify that this is true in our implementation of the monochromatic theory [\Eq{eq:mono}], we studied the thermal radiation of a spherical nearly black body. This object is a sphere with \SI{10}{\micro\meter} radius made of a material that has $\epsilon = 1 + 0.2i$ as shown in Fig.~\ref{fig:blackbody}(a). The reflection coefficient at the sphere-air interface is small (Fresnel reflectance ca. 1\% at normal incidence) while the absorption inside the sphere is strong. Consequently, for wavelengths much shorter than the radius of the sphere, the sphere behaves like a nearly black body with an absorption cross section equal to its geometric cross section [green solid and black dotted lines in Fig.~\ref{fig:blackbody}(b)]. Looking at the thermal radiation spectrum, we get the expected result: For the monochromatic theory, the thermal radiation spectrum of the sphere (red solid curve in Fig.~\ref{fig:blackbody}(c)) overlaps with the Planck spectrum (black dashed curve) at these wavelengths.  
To finish, we consider \cite[Eq.~(B7)]{Mazo2025}:
\begin{equation}
	\label{eq:aiai}
	\exval{|\bra{f}ks\rangle|^2}=\sum_\lambda\exval{|f_\lambda(k)|^2}\sum_{jm}|\vec{s}_{jm\lambda}(k)|^2,
\end{equation}
particularized for the Planckian with 
\begin{equation}
\exval{|f_\lambda(k)|^2}=\exval{|\psi_+(k)|^2} = \frac{\cz}{2\pi k\left(\exp\left(\frac{\hbar \cz k }{k_B \text{T}}\right)-1\right)},
\end{equation}
with which we obtain the final result from \Eq{eq:aiai}:
\begin{equation}
	\begin{split}
	\exval{|f_s(k)|^2}&=\frac{\cz}{2\pi k\left(\exp\left(\frac{\hbar \cz k }{k_B \text{T}}\right)-1\right)}\sum_{jm\lambda}|\vec{s}_{jm\lambda}(k)|^2
	\\
	&=\frac{\cz}{2\pi k\left(\exp\left(\frac{\hbar \cz k }{k_B \text{T}}\right)-1\right)}.
\end{split}
\end{equation}

\bibliography{polychromatic}
\end{document}

%% file: new_preamble_definitions.tex
\usepackage[uline]{hhtensor}
\usepackage{mathrsfs}
\usepackage{braket}
\usepackage{hyperref}
\usepackage{mathtools}
\usepackage{siunitx}
\usepackage{xcolor}
\usepackage{textgreek}
\usepackage{framed}
\usepackage{empheq}

\newcommand\ff[1]{\mathrm{f}_{#1}(\pp)}

\newcommand\ffe[1]{f_{#1}(\pp)}
\newcommand\ffestar[1]{f^*_{#1}(\pp)}
\newcommand\FFe[1]{f_{jm#1}(k)}
\newcommand\FFestar[1]{f^*_{jm#1}(k)}

\newcommand*{\hbaraux}[2]{\sbox0{\mathsurround=0pt$#1\mkern-1mu\mathchar'26$}\mkern-1mu\lower.07\ht0\box0\mkern-8mu}
\renewcommand*{\hbar}{{\mathpalette\hbaraux\relax\mathrm{h}}}

\newcommand\impliesduetoeq[1]{\stackrel{Eq.\ (\ref{#1})}{\implies}}

\renewcommand\Im[1]{\text{Im}\left\{#1\right\}}

\newcommand\Eq[1]{\text{Eq.~(\ref{#1})}}

\newcommand\rr{\mathbf{r}}

\newcommand\pp{\mathbf{k}}

\newcommand\phat{\mathbf{\hat{\pp}}}

\newcommand\Ert{\mathbf{E}(\rr,t)}

\newcommand\intdpinv{\int_{\mathbb{R}^3} \frac{\text{d}^3 \pp}{k}\text{ }}

\newcommand\zerovec{\mathbf{0}}

\newcommand\ii{\mathrm{i}}

\newcommand\epsz{\text{\textepsilon}_{\text{0}}}
\newcommand\cz{\mathrm{c_0}}

\newcommand\rhat{\mathbf{\hat{r}}}

\newcommand{\mybraket}[2]{\langle#1|#2\rangle}

\newcommand\Id{\mathrm{I}}
\newcommand\op[1]{\mathrm{#1}}

\newcommand\exval[1]{\mathbb{E}\left\{#1\right\}}
\newcommand\barlambda{\bar{\lambda}}

\newcommand\qq{\mathbf{q}}

\newcommand\ketbathe{\ket{\Psi_\text{bath}}}
\newcommand\trace[1]{\text{trace}\left\{#1\right\}}

\newcommand\intdkmeasure{\int_{>0}^\infty {\mathrm{d}k}k}

\newcommand\jbar{\bar{j}}

\newcommand\mbar{\bar{m}}

\newcommand{\sandwich}[3]{\langle#1|#2|#3\rangle}
\newcommand\inout{\text{in/out}}
\providecommand\M{\mathbb{M}}

\providecommand\tauabs{\tau^{\text{abs}}}
\providecommand\tausabs{\tau^{\text{abs}}_s}

\providecommand\wsp{\omega_{sp}}
\providecommand\wspstar{\omega^*_{sp}}
\providecommand\wspre{\omega_{sp}^\text{re}}
\providecommand\wspim{\omega_{sp}^\text{im}}

\providecommand\gammaspem{\gamma_{sp}^\text{em}}

\providecommand\gammaabs{\gamma^\text{abs}}
\providecommand\gammaem{\gamma^\text{em}}
\providecommand\Hsp{\text{H}_{sp}}
\providecommand\Hspbar{\text{H}_{s\bar{p}}}

\providecommand\ketsabs{\ket{ks}^{\text{abs}}}
\providecommand\ketsem{\ket{ks}^{\text{em}}}

\providecommand\brasabs{{}^{\text{abs}}\hspace{-0.1cm}\bra{sk}}
\providecommand\ketspem{\ket{sp}^\text{em}}
\providecommand\braspem{{}^{\text{em}}\bra{ps}}